\documentclass[aps,prl,twocolumn,showpacs]{revtex4}

\usepackage{amsmath}

\begin{document}

\input psfig.sty

\title{Even-odd effects in magnetoresistance of ferromagnetic domain walls}
\author{M.\ Dzero
\footnote{Also at: Department of Physics, Florida State University.
Electronic address: dzero@magnet.fsu.edu}$^{,1}$,
L.\ P.\ Gor'kov$^{1,2}$, A.\ K.\ Zvezdin$^{1,3}$ and K. A. Zvezdin$^{3}$}
\affiliation{$^1$National High Magnetic Field Laboratory, Florida State
University, Tallahassee, FL 32310, USA}
\affiliation{$^2$L.D. Landau Institute for Theoretical Physics, Russian Academy
of Sciences, 117334 Moscow, Russia}
\affiliation
{$^3$ Institute of General Physics, Russian Academy of Sciences, 117942 Moscow, Russia}

\begin{abstract}
Difference in density of states for the 
spin's majority and minority bands in a ferromagnet changes 
the electrostatic potential along the domains, introducing the discontinuities
of the potential at domain boundaries. The value of discontinuity oscillates with 
number of domains. Discontinuity depends on the positions of 
domain walls, their motion or collapse of domain walls in applied magnetic field.
Large values of magnetoresistance are explained in terms of spin-accumulation.
We suggest  a new type of domain walls in nanowires of itinerant ferromagnets, in which
the magnetization vector changes without rotation. Absence of transverse magnetization components 
allows considerable spin accumulation assuming the spin relaxation length, $L_S$, 
is large enough.
\end{abstract}
\maketitle

As it was demonstrated first in \cite{Johnson,Sohn}, spin accumulation effects in the
presence of a current flowing between ferromagnetic and normal metals 
may result in a considerable contribution to a contact's resistivity.
This phenomenon is the key element for the GMR devices with the CPP (current perpendicular to
the plane) geometry. The CPP-GMR effect has been studied theoretically for the spin-valve 
systems for a simple triple 
layer and multi-layered structures in the pioneering work \cite{Valet}. The subsequent experiments
(see e.g. \cite{Pratt,Piraux,Dubois}) were in excellent agreement with the predictions made in \cite{Valet}
concerning the dependence of the resistivity on the width of magnetic and non-magnetic components
and the role played by spin-relaxation mechanisms. 
Most recently it was discovered that nanocontacts \cite{Garcia} and domain walls in magnetic nanowires
\cite{Ebels} possess significant magnetoresistance.

Realization of different experimental configurations allows determination of parameters present in the expressions
in Ref. [\onlinecite{Valet}] such as the values of resistivity for each of the GMR components 
and what is most important the spin
relaxation length, $L_S$, characterizing the width of a non-equilibrium distribution of spins near the
contacts \cite{Piraux,Dubois}.

The formulas in Ref. [\onlinecite{Valet}], however, have been derived in assumption that while the conductivities
of the majority and minority spins are different, the corresponding densities of states (DOS), $g_{\alpha}$, 
remain equal. This assumption is not realistic. In the presentation below we address this issue to demonstrate that
taking difference in the DOS into account, the changes in the expressions for the distribution of the
electrostatic potential lead to some new observable effects. 
In Ref. [\onlinecite{Garcia,Ebels}] it was speculated that the pronounced GMR effects are caused
by the significant role of spin accumulation. 
We suggest, as we believe, for the first time,
that large magnetoresistance observed in Ref. [\onlinecite{Ebels}] is due to the non-rotational character of the 
domain walls \cite{Zhirnov} which are possible in itinerant ferromagnets \cite{Dzero}.

Following \cite{Valet}, we re-write the expression for the current $j_{\alpha} = eD_{\alpha}n{'}_{\alpha} -
\sigma_{\alpha}U'(x)$, ($U(x)$ - the electrostatic potential, the index, $\alpha=\pm$, stands for the majority and
minority spins correspondingly), into the form:
\begin{equation}
{\mbox{\boldmath $j$}}_{\alpha} = \frac{\sigma_{\alpha}}{e}{\mbox{\boldmath $\nabla$}}\mu_{\alpha},
\label{eq1}
\end{equation}
where
\begin{equation}
\mu_{\alpha} = g^{-1}_{\alpha}n_{\alpha} - eU(x)
\label{eq2}
\end{equation}
is a non-equilibrium (in the presence of a total current, ${\mbox{\boldmath $J$}}$) electrochemical
potential for each spin component, and the relation $D = \sigma/(e^2g)$ is used in (\ref{eq1}, \ref{eq2}).

The equations for the current components:
\begin{equation}
\text{div}~{\mbox{\boldmath $j$}}_{\alpha} = \frac{en_{\alpha}}{\tau_S}
\label{eq3}
\end{equation}
together with (\ref{eq2}) and the electro-neutrality condition:
\begin{equation}
\sum\limits_{\alpha}^{}n_{\alpha} = n_{+} + n_{-} = 0
\label{eq4}
\end{equation}
present the complete system of equations for each side of an interface (in (\ref{eq3}) $\tau_S$ is a
spin relaxation time). 

To simplify the analysis, we at first assume the ballistic regime for the interface, i.e.
the width of the corresponding domain wall is small (contributions due to the spin
scattering inside the barrier will be discussed at the end of the paper).  
Correspondingly $\mu_{\alpha}$ and $j_{\alpha}$ are taken continuous at the boundary.

In convenient notations of Ref. [\onlinecite{Rashba}] with the help of (\ref{eq4}) we obtain
for $\mu_S = \mu_{+} - \mu_{-}$:
\begin{equation}
\mu_S = \left(g^{-1}_{+} + g^{-1}_{-}\right)n_{+}.
\label{eq5}
\end{equation}

Subtracting one of Eqs. (\ref{eq3}) from another and making use of (\ref{eq5}), one obtains
the equation for the distribution of $\mu_S$ in the ``bulk'' on each side of the contact:
\begin{equation}
\mu{''}_S = \frac{\mu_S(x)}{L^2_S},
\label{eq6}
\end{equation}
where
\begin{equation}
L^{-2}_S = \frac{e^2}{\tau_S}\left(\frac{\sigma^{-1}_{+} + \sigma^{-1}_{-}}
{g^{-1}_{+} + g^{-1}_{-}}\right).
\label{eq7}
\end{equation}
Equation (\ref{eq6}) coincides in its form with Eq. (14) from
Ref. [\onlinecite{Valet}]. For our purposes it is convenient to
re-write (\ref{eq6}) as:
\begin{equation}
\mu{''}_S - \mu_S/L^2_S = \delta(x)\Delta(\mu{'}_S).
\tag{6'}
\label{eq6p}
\end{equation}
Equation (\ref{eq6}) applies now everywhere in the sample. In (\ref{eq6p}), $\delta(x)$ stands for the
Dirac delta-function at the boundary ($x=0$) (equation ((\ref{eq6p}) is valid
when the boundary between the domains is abrupt). While $\mu_S$ is continuous, $\mu{'}_S$ has a jump at the
interface:
\begin{equation}
\Delta(\mu{'}_S) = \mu{'}_S(+0) - \mu{'}_S(-0).
\label{eq8}
\end{equation}
The jump can be expressed in terms of the total current, $J$. To do so, we write down
the expressions for $J$, spin current $j_S = j_{+} - j_{-}$, to obtain with the help
of (\ref{eq4},\ref{eq5}):
\begin{equation}
\begin{split}
J = \frac{1}{e}\left(\frac{\sigma_{+}g^{-1}_{+} - \sigma_{-}g^{-1}_{-}}{g^{-1}_+ + g^{-1}_{-}}\right)
\mu'_{S}(x) - (\sigma_{+}+\sigma_{-})U'(x),\\
j_S = \frac{1}{e}\left(\frac{\sigma_{+}g^{-1}_{+} + \sigma_{-}g^{-1}_{-}}{g^{-1}_+ + g^{-1}_{-}}\right)
\mu'_{S}(x) - (\sigma_{+}-\sigma_{-})U'(x).
\end{split}
\label{eq9}
\end{equation}
We specify that \emph{on the left} from the domain wall spins up belong to the majority band.
Magnetization changes sign across the wall and so does the band's occupation so that 
one has to interchange between $\sigma_{\pm},
g^{-1}_{\pm}$ in (\ref{eq9}). Since the currents are continuous, for the jump of electrochemical potential
gradient, $\Delta(\mu'_S)$, we get:
\begin{equation}
\Delta(\mu'_S) = e\left(\frac{\sigma_{+}-\sigma_{-}}{\sigma_{+}\sigma_{-}}\right)J\equiv{C}.
\label{eq10}
\end{equation}
For a multilayer system, Eq. (\ref{eq6p}) reads:
\begin{equation}
\mu{''}_S - \frac{\mu_S}{L^2_S} = C\sum\limits_{i=1}^{N}(-1)^{i+1}\delta(x-x_i), (i=1,...,N).
\label{eq11}
\end{equation}
It is also assumed below that the width of the very left and very
right banks are larger then $L_S$. The solution of (\ref{eq11})
is a superposition of solutions for a single domain wall:
\begin{equation}
\mu_S(x) = -\frac{CL_S}{2}\sum\limits_{j=1}^N{(-1)^j}{e}^{-\mid{x-x_j}\mid/L_S}.
\label{eq12}
\end{equation}

We consider first the behavior of the electrostatic potential, $U(x)$, close to a single domain 
wall. With the help of Eq. (\ref{eq5}) one can write:
\begin{equation}
\mu_{+} + \mu_{-} = - \left(\frac{g_{+}-g_{-}}{g_{+}+g_{-}}\right)\cdot\mu_S(x) - 2eU(x).
\label{eq13}
\end{equation}
Making use of continuity of the electrochemical potential at the boundary, one immediately sees
from (\ref{eq13}) that potential $U(x)$ is discontinuous with the jump $\Delta{U_1}(x_i) = 
U(x=x_i,R) - U(x=x_i,L)$ at the $i$th domain wall:
\begin{equation}
\Delta{U_1}(x_i) = \frac{(-1)^i}{e}\cdot\left(\frac{g_{+}-g_{-}}{g_{+}+g_{-}}\right)\cdot\mu_S(x_i).
\label{eq14}
\end{equation}
The spatial distribution of $U(x)$ can be found by integrating the first equation (\ref{eq9}) 
along each side of the domains:
\begin{equation}
\begin{split}
&U'(x) = -\frac{J}{\sigma_{+}+\sigma_{-}} + 
\frac{1}{2e}\left[\left(\frac{\sigma_{+}-\sigma_{-}}{\sigma_{+}+\sigma_{-}}\right)\right.\\
&\quad\quad-\left.\left(\frac{g_{+}-g_{-}}{g_{+}+g_{-}}\right)\right]\cdot\mu{'}_S.
\end{split}
\label{eq15}
\end{equation}

The total potential drop across an isolated single domain wall in the chosen geometry is the sum
of two terms:
\begin{equation}
\begin{split}
\Delta{U} = \Delta{U}_1 + \frac{1}{e}\left(
\frac{\sigma_{+}g^{-1}_{+} - \sigma_{-}g^{-1}_{-}}{g^{-1}_{+} + g^{-1}_{-}}
\right)\times\\
~2\int\limits_{0}^{\infty}\mu^{'}_S(x){dx} =
\frac{L_S}{2\sigma_{+}\sigma_{-}}\cdot\frac{(\sigma_{+} - \sigma_{-})^2}{(\sigma_{+} + \sigma_{-})}J
\end{split}
\label{eq16}
\end{equation}
(the second term in (\ref{eq16}) comes about from the currents distribution (\ref{eq12}) on
the distances of the order of $L_S$ on the two sides of the wall). In the following notations
\begin{equation}
\sigma_{\pm}=\frac{\sigma}{2}(1\pm\beta), ~g_{\pm}=\frac{g}{2}(1\pm\delta),
\nonumber
\end{equation}
one obtains:
\begin{equation}
\Delta{R} = \frac{\Delta{U}}{J} = 2\frac{\beta^2{L_S}}{\sigma(1-\beta^2)},
\label{eq17}
\end{equation}
i.e. $g_{\pm}$ drop out from the \emph{total} magnetoresistance.
Equations (\ref{eq16}-\ref{eq17}) coincide with the results obtained in \cite{Valet}.
The differences in the DOS for the minority and majority spins lead to an appearance of the
discontinuities (\ref{eq14}) and changes in the dependence of the electrostatic potential
$U(x)$ (see (\ref{eq15})) along the domains. 

To make a numerical estimate for the total potential drop we will use the data from 
Ref. [\onlinecite{Ebels}] obtained for Co nanowires: $L_S\simeq{60}$nm, 
$\rho\simeq{1.3\cdot{10}^{-5}}~\Omega\cdot{cm}$. The typical values of
$\delta$ are of the order of ${0.4}\div{0.5}$~\cite{Piraux, Dubois}, while 
$\beta$ is of the order of $0.5\div{0.7}$~\cite{Janak}. After substituting these values in (\ref{eq17}), 
the resistance drop $\Delta{R}$ per unit area is $5\cdot{10^{-11}}~\Omega$, 
or $\Delta{R}\simeq{112}~\Omega$ for the geometry used in \cite{Ebels}.
The value of the potential drop at domain boundary is approximately the same as the value
of the total potential drop in (\ref{eq16}): 
\begin{equation}
\Delta{U}_1/\Delta{U} = \delta/\beta. 
\tag{17'}
\label{eq17p}
\end{equation}
We would like to emphasize, that the ratio in (\ref{eq17p}) may also be negative.

As an example, let us consider in some more details, the drop of the potential, $\Delta{U}(x_i)$,
across the very left domain for a system of $N$ walls (we also take $x_1=0$). 
From (\ref{eq12}) and (\ref{eq14}) we have
\begin{equation}
\Delta{U}_N = \frac{1}{2}
\frac{(g_{+}-g_{-})\cdot(\sigma_{+}-\sigma_{-})}{\sigma_{+}\sigma_{-}(g_{+}+g_{-})}JL_SS_N,
\label{eq18}
\end{equation}
where 
\begin{equation}
S_N = \sum\limits_{j=1}^N~(-1)^j{e}^{-x_j/L_S}.
\tag{18'}
\label{eq18p}
\end{equation}
For the $N$ walls the value of $\Delta{U}_N$ 
shows an interesting ``even-odd'' effect:
\begin{equation}
\begin{split}
\left(\delta{U}\right)_N\equiv\left(\Delta{U}_{N+1} - \Delta{U}_{N}\right) = \\
 = \frac{(-1)^N}{2}\left(\frac{g_{+}-g_{-}}{g_{+}+g_{-}}\right)\cdot
\left(\frac{\sigma_{+}-\sigma_{-}}{\sigma_{+}\sigma_{-}}\right)J{e}^{-Nd/L_S}.
\end{split}
\label{eq19}
\end{equation}
Below in Fig. \ref{Fig1} we plot $\left(\delta{R}\right)_N/\Delta{R}=\left(\delta{U}\right)_N/J\Delta{R}$ as
a function of number of the domain walls. The ``even-odd'' effect is well-pronounced at 
$Nd\sim{L_S}$, where $d$ is a size of domain. 
\begin{figure}
\hspace{0cm}
\centerline{\psfig{file=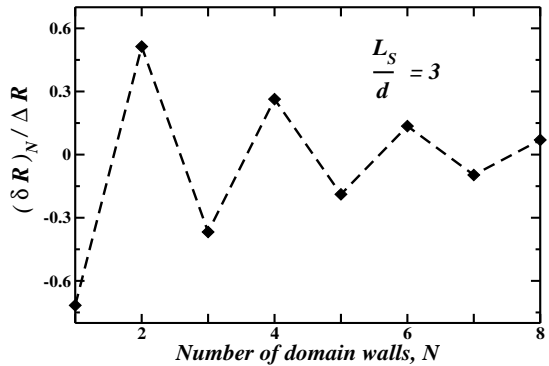,height=6cm,width=8cm,angle=-90}}
\caption{Ratio $\left(\delta{R}\right)_N/\Delta{R}$ is plotted as a function of N (\ref{eq19}).
The distance between the domain walls is estimated using the formula 
$d\simeq\sqrt{d_wl_{Co}}$ ($d_w\simeq{10}$nm is a width of the domain wall, 
$l_{Co}\simeq{40}$nm is the diameter of a wire \cite{Ebels}).} 
\label{Fig1}
\end{figure}

\begin{figure}
\hspace{0cm}
\centerline{\psfig{file=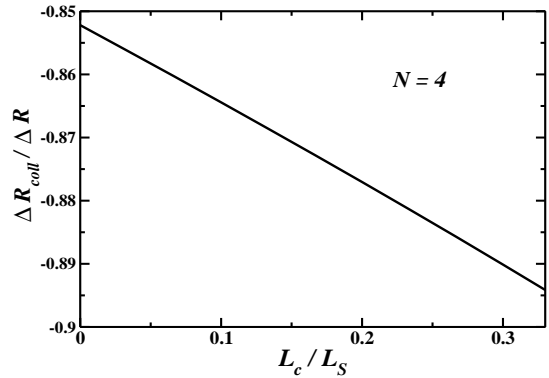,height=6cm,width=8cm,angle=-90}}
\caption{Ratio $\Delta{R_{coll}}/\Delta{R}$ is plotted as a function of dimensionless parameter 
$l_c=L_c/L_S$ showing the resistivity drop in the sample corresponding to the collapse of two domain walls.
For simplicity, we assumed that there have been four domain walls before an external magnetic 
field was applied.} 
\label{Fig2}
\end{figure}

Another interesting feature is that changes in Eq. (\ref{eq18}) could 
follow a motion of a single domain wall, say at $x=x_i$,
through its contribution $(-1)^i{e}^{-x_i/L_S}$ into (\ref{eq18p}), 
or even a collapse of a domain caused by the applied magnetic field 
(such a collapse has been experimentally seen in Ref. [\onlinecite{Ebels}]). In order to explicitly
demonstrate this effect, we first introduce the following notations:
\begin{equation}
\frac{\Delta{R_{coll}}}{\Delta{R}} = \sum\limits_{j=1}^{i}~(-1)^j{e}^{-x_j/L_S} +
\sum\limits_{j=i+1}^{N}~(-1)^j{e}^{-x_j/L_S}.
\nonumber
\end{equation}
The result of our calculations of $\Delta{R_{coll}}/\Delta{R}$ as a function of the distance between 
two neighboring domains $L_c = x_{i+1} - x_{i}$ is plotted in Fig. \ref{Fig2}. Experimentally, 
the motion of a domain wall can be detected by the STM technique.

Now we would like to briefly discuss the change in our results in case when
one takes the finite width of the domain wall into account.
In particular, we consider the importance of depolarizing effects in the Bloch or Ne\'{e}l
type of domain wall for the results above. Since the width of the wall is usually much
smaller then the spin diffusion length, $L_S$, the arguments that
lead us to results given by (\ref{eq14}-\ref{eq16}) still hold.
The only modification would come from the change in the boundary
conditions for the spin current.

Electrons going through the Bloch or Ne\'{e}l domain wall
loose part of its polarization because the transverse
component of magnetization inside the wall creates a torque which
causes spin's re-orientation. Obviously this process will reduce
the spin accumulation. If $d_w$ is the width of a domain wall, spins of
electrons traveling through the wall are rotated in the transverse component
of the exchange field, $H_{exch}$, in the wall by an angle $\vartheta_{\alpha} (\alpha=\uparrow\downarrow)$:
\begin{equation}
\vartheta_{\alpha} = (\pm)\frac{\mu_BH_{exch}d_w}{\hbar{v}_{\alpha}}.
\label{eq20}
\end{equation}
From (\ref{eq20}) one obtains:
\begin{equation}
j_S(x=+0) - j_S(x=-0) = 2en_{+}v
\label{eq21}
\end{equation}
with $v$ given by:
\begin{equation}
v = \frac{2\mu_BH_{exch}d_w}{\hbar}
\tag{21'}
\label{eq21p}
\end{equation}
and after some algebra with help of (\ref{eq5}):
\begin{equation}
j_S(x=+0) - j_S(x=-0) = {2ev}\frac{g_{+}g_{-}}{g_{+} + g_{-}}\mu_S(0).
\label{eq22}
\end{equation}
Taking into account (\ref{eq22}), the expression for the jump of $\mu'_S$ at the interface
is:
\begin{equation}
\begin{split}
&\Delta(\mu'_S) = e\left(\frac{\sigma_{+}-\sigma_{-}}{\sigma_{+}\sigma_{-}}\right)J ~ \\
&~ + ~\frac{e}{2}\left[\frac{2evg_{+}g_{-}(\sigma_{+}+\sigma_{-})}
{(\sigma_{+}\sigma_{-})(g_{+}+g_{-})}\right]\cdot\mu_S(0).
\end{split}
\label{eq23}
\end{equation}
As a result, the expression for the potential drop $\Delta{U}$ across an isolated domain 
wall acquires the form:
\begin{equation}
\begin{split}
\Delta{U} = \frac{\beta^2}{(1-{\beta}^2)}\cdot\frac{2L_S{\sigma^{-1}}}{1+\alpha}J, \\
\alpha = \frac{e^2vgL_S{\sigma^{-1}}}{2}\frac{1-{\delta}^2}{1-{\beta}^2}.
\end{split}
\label{eq24}
\end{equation}

To estimate the value of $\alpha$, we first re-write the second expression in (\ref{eq24}) as:
\begin{equation}
\alpha\simeq 3\left(\frac{v}{v_F}\right)\cdot\frac{L_S}{l},
\label{eq25}
\end{equation}
where $l$ is a mean free path and $v_F$ is the Fermi velocity. Using the data, provided in \cite{Ebels}, 
we have: $\mu_BH_{exch}\sim{0.1}$eV, $d_w\simeq{10}$nm, $L_S/l\simeq{10}$, 
$v\simeq{4.5}\cdot{10}^{7}$cm/s, $v_F\simeq{10^8}$cm/s. 
Thus our estimates provide $\alpha\simeq{14}$, which according to (\ref{eq23})
significantly reduces the spin accumulation effect. This result can also be explained 
slightly differently: from (\ref{eq20}) we estimate the depolarization angle $\vartheta\simeq{1}$, 
i.e. at crossing the Bloch or Ne\'{e}l type of domain wall the electronic spins would 
adiabatically follow magnetization. To reconcile the above estimates with 
the significant magnetoresistance experimentally observed in Ref. [\onlinecite{Piraux,Ebels}], 
we suggest that the domain walls in these experiments were neither of the Bloch nor Ne\'{e}l type. 
Instead, in itinerant ferromagnets another type of domain walls, ``linear walls'', is realized \cite{Dzero}.
In a linear wall the direction of magnetization does not change, while its absolute
value goes through zero inside the wall. Theoretically, ``linear'' domain walls were considered 
in \cite{Zhirnov} for local spin ferromagnetic systems at temperatures $T$ close to critical temperature $T_{Curie}$. 

To summarize, we have shown that taking into account difference in the density of states between the minority
and majority spin bands drastically changes the distribution of the electrostatic potential along the domains. 
The discontinuities of the potential across each domain wall are of particular interest. The jumps
in the values of potential can be measured directly by the STM technique. Jumps possess such characteristic
features as the ``even-odd'' effects counting the total number of domain walls in magnetic nanowire in the 
presence of an external magnetic field. Our results can be directly extended to the GMR structures 
consisting of F/N layers which were studied theoretically in \cite{Valet} in the approximation of equal 
density of states of majority and minority bands. To ascribe large values of magnetoresistance observed
in \cite{Piraux,Ebels} to spin accumulation effects, it was also necessary to suggest that in nanowires
made of itinerant ferromagnets the domain walls are of a linear type in which magnetization changes without
creating a perpendicular component that would revert spins of polarized electrons. 

The work of M.D. was supported by DARPA through the Naval Research Laboratory
Grant No. N00173-00-1-6005, L.P.G. by the NHMFL through the
NSF cooperative agreement DMR-9527035 and the State of Florida,
A.K.Z. through the NISVP Supplement to the core NHMFL Grant 
No. 5024-599-22 and the State of Florida, and K. A. Z. by the RFBR grant No. 
02-02-17389.

\end{document}